\documentclass[trackchanges,twocolumn]{aastex701}
\usepackage{siunitx}
\usepackage{amsmath}
\newif\ifref
\reffalse
\definecolor{darkred}{rgb}{0.75, 0, 0}
\newcommand{\mb}[1]{\ifref\textcolor{darkred}{#1}\else #1\fi}

\begin{document}  

\title{Predicting the Kinematics of the Cold Circumgalactic Medium from its Morphology using Convolutional Neural Networks}

\author[0009-0008-3897-4149]{Connor Jennings}
\affiliation{Astronomy Department, Yale University, 219 Prospect St, New Haven, CT 06511, USA}
\email[show]{connor.jennings@yale.edu}  

\author[0000-0003-4456-4863]{Earl P.\ Bellinger}
\affiliation{Astronomy Department, Yale University, 219 Prospect St, New Haven, CT 06511, USA}
\email{earl.bellinger@yale.edu}

\author[0000-0002-7075-9931]{Imad Pasha} 
\affiliation{Astronomy Department, Yale University, 219 Prospect St, New Haven, CT 06511, USA}
\affiliation{Dragonfly Focused Research Organization, 150 Washington Avenue, Suite 201, Santa Fe, NM 87501, USA}
\email{imad.pasha@yale.edu}

\author[0000-0002-8282-9888]{Pieter van Dokkum}
\affiliation{Astronomy Department, Yale University, 219 Prospect St, New Haven, CT 06511, USA}
\affiliation{Dragonfly Focused Research Organization, 150 Washington Avenue, Suite 201, Santa Fe, NM 87501, USA}
\email{pieter.vandokkum@yale.edu}

\author[0000-0003-0965-605X]{Pratik J. Gandhi}
\affiliation{Astronomy Department, Yale University, 219 Prospect St, New Haven, CT 06511, USA}
\email{pratik.gandhi@yale.edu}

\begin{abstract}

We present a novel approach to predicting plane-of-sky velocities of cold gas clouds in the circumgalactic medium (CGM) of galaxies. The method uses a convolutional neural network (CNN) trained on simulated emission maps derived from the TNG50 cosmological simulation, with forward modeled noise properties consistent with upcoming observational facilities. Using 182 Milky Way/Andromeda analog galaxies, we generate emission maps in H$\alpha$ using \texttt{Cloudy} models, as well as line-of-sight averaged 2D velocity maps. Using a UNet architecture, we train the CNN to take emission maps as input and return plane-of-sky velocity maps as output, which cannot be observationally constrained using traditional methods. Qualitatively, the model is generally able to infer the true overall flow direction.  We quantify the effects of Gaussian noise on the network's training and predictive power. At depths expected to be probed by forthcoming telescopes such as MOTHRA, the network has a typical RMS error for the plane-of-sky velocity direction of $0.3-0.5 v_{vir}$. This implies that 2D emission maps of sufficient depths will be able to estimate two additional phase space dimensions of cold CGM gas, enabling targeted followup and a better understanding of overall CGM flows.

\end{abstract}


\section{Introduction}

The circumgalactic medium (CGM) is the volume of gas lying in between a galaxy's interstellar medium (ISM) and the pristine intergalactic medium (IGM)  \citep[see reviews by][]{Tumlinson2017_CGMreview,Faucher2023_review,Chen2026_review}. The CGM is a reservoir which processes both inflowing gas passing from the IGM to the ISM to fuel star formation or maintain active galactic nuclei (AGN) activity within the galaxy, as well as outflowing gas ejected from the galaxy through stellar or AGN feedback. Therefore, understanding the nature of CGM gas---its temperature, its chemical composition, where it is, and where it is going---is key to understanding the evolution of galaxies.

The CGM is multiphase, with multiple temperatures of gas in rough pressure equilibrium \citep{Hui1997_EOS_of_IGM,Putman2012_Gaseous_Galaxy_Halos_review}. The majority of the CGM's volume is filled by gas in the hot ($>10^6K$) phase, which is interspersed by much denser cold ($\sim10^4K$) phase gas. A warm phase forms as a boundry layer between the cold and hot phases. Gas can be heated into the hot phase via stellar feedback \citep{Conroy2015_Stellar_heating}, AGN feedback \citep{Fabian2012_AGN_Feedback_Observation}, as well as from virial shock heating as gas flows into the CGM from the IGM \citep{Dekel2006_Shock_Heating_and_Cold_flows,Stewart2011_Virial_shock}. Gas can cool out of the hot phase and into the cool phase through processes such as the precipitation model \citep{Voit2015_precipitation,Voit2017_Percipitation} or a condensation cascade \citep{Gaspari2018_CondensationCascade}, leading it to fall towards its host galaxy. The quenching of massive galaxies requires some feedback mechanism to heat the ISM and CGM gas faster than it can cool and form new stars. Cosmological simulations broadly recover the stellar masses and star formation rates of galaxies \citep{Wright2024_BaryonCycleSimulations}, but there is large disagreement between different simulations as to the temperature and amount of CGM gas around galaxies in a range of masses \citep{Wright2024_BaryonCycleSimulations,Medlock2025_CAMELSBaryoneCycle} with differences between simulations largely being attributed to different subgrid prescriptions for baryonic feedback. 

The most extensive observational studies of the CGM to date have utilized quasi-stellar object (QSO) absorption spectra \citep[e.g.,][]{Schroetter2016_MEGAFLOW1_absorption,Chen2020_CUBS1_absorption,Chen2023_CUBS_turbulence,Huang2021_MgII_absorption_lowZ,Dutta2025_OIV_absorption,Chang2025_DESI_year1}. Depending on the lines observed, absorption spectra can constrain the amount, temperature, metal enrichment, and non-thermal broadening of both hot and cold CGM phases out to large distances from the host galaxy, with detection limits scaling as $n^{-1}$ with gas density. However, these studies can only measure gas that happens to fall along the line of sight to the QSO, limiting most studies to a single pencil-beam measurement. Some morphological studies can still be performed by compiling angles of absorption off of edge-on galaxies \cite[see][where higher metal abundances are found along the minor axis, consistent with enrichment via bipolar AGN outflows]{Kacprzak2015_azimuthal_dependance_absorption} or selecting for rare, multiply-lensed QSOs which can net 2 or 4 absorption measurements around a single galaxy \citep{Chen2014_multiply_lensed_QSO_absorption}, but neither of these methods can resolve gas to the $\sim$kpc cloud scale expected from simulations \citep{Ramesh2023_TNG50CGM1}.

Another method of CGM observation is through emission. Cold gas, ionized by the UV background \citep{Cantalupo2005_Lya_UV_background,Faucher2009_TNG_UVbackground,Faucher2020_UVbackground} can produce detectable emission lines in Lyman~$\alpha$ \citep{Gallego2018_MUSE_Lya_stack}, H$\alpha$ \citep{Fumagalli2017_Ha_UVbackground}, and metal lines \citep{Zabl2021_MEGAFLOW_MgII_emission}. Compared to absorption, emission measurements have a steep $n^{-2}$ density dependence \citep{Dijkstra2017_Physics_of_Lya_transfer}. Still, studies using KCWI \citep{Chen2021_KBSS_KCWI_La_emission_z23}, MUSE \citep{Leclercq2017_MUSE_La_emission}, and CHaS \citep{Melso2024_CHaS_Ha_emission} have detected extended emission out into the CGM, allowing the mapping of smaller scale structures than can be observed via absorption lines.

In this paper, we specifically compare mock H$\alpha$ emission from simulated Milky Way (MW)-mass galaxies to the expected capabilities of the upcoming MOTHRA telescope. The completed MOTHRA array will have 1140 telephoto lenses with a combined effective diameter of 4.8 meters and a $2^\circ \times 3^\circ$ field of view. The array is designed with the goal of seeing down to a surface brightness of $10^{-20.5}\text{ergs}/\text{s}/\text{cm}^2/\text{arcsec}^2$ in H$\alpha$, NII, and OIII emission, with the ultimate aim of imaging the cosmic web (private communication). This depth will allow the observation of major cold gas streams in the CGM around local, $z<0.01$, galaxies \citep{Lokhorst2019_MOTHRA_detectability,Pasha2026}. 

Once the full 2D spatial distributions of cold CGM gas are resolved, the next question will be to predict where the gas is going. Plane-of-sky velocities are difficult to measure directly in extragalactic astronomy, but the cold gas in the CGM obeys different dynamics than collisionless stars, which might allow better predictions. For instance, gas stripped off of jellyfish galaxies by ram pressure can be used to infer trajectories relative to the intracluster medium \citep{McPartland2016_Jellyfish,delosRios2021_ROGER}, and the trajectory of the gas itself can be inferred using inclination estimates and moment velocity maps \citep{Souchereau2025}. \cite{Gupta2024_TIPSY} were able to reconstruct the trajectories of infalling streams onto young stellar objects (YSOs). However, these methods rely on gas originating all from the same galaxy or point in the ISM, and use high resolution velocity maps. Cold CGM gas does not originate from a single position, so the methods used for Jellyfish galaxies and streamers onto YSOs cannot be copied directly, but the principle that cold gas moving through a medium encodes information about its trajectory in its morphology still holds even without resolving the line-of-sight velocity dimension. 

For cold CGM clouds passing through the volume filling hot phase there exists some non-injective, non-linear mapping from images of emission to maps of velocity.
For tasks involving inferring properties based on images, computer vision utilizing deep neural networks has shown promise in astrophysical applications and has been used extensively to classify galaxies based on morphology \citep{Khalifa2017_DeepGalaxy_classifier,Tohill2021_CNN_classifier_CANDLES,Rosito2023_ClusteringDimensionalityReductionGalaxies}, as well as on simulated data to predict global galaxy properties, such as stellar mass, through CGM emission \citep{Gluck2024_21cmCNN}.

\mb{While there are many potential observables and values of interest that can characterize CGMs, in this paper we focus on recovering kinematics from H$\alpha$ emission morphology in order to characterize what information is encoded in that morphology.} We use a convolutional neural network (CNN) to predict the plane-of-sky velocities of CGM gas. Using a \texttt{UNet} architecture, we encode \mb{maps of H$\alpha$ surface brightness} and decode to recover a 2D velocity vector for each pixel in our image. \texttt{UNet} was originally developed for medical imaging \citep{Ronneberger2015_UNet}, and takes a 2D array as both its input and output. For our implementation, our input is a synthetic map of H$\alpha$ emission from simulated galaxies, and our outputs are two 2D arrays corresponding to the two components of the line-of-sight averaged plane-of-sky velocity field. We describe our dataset and model training in Section \ref{sec:Methods}. We give an overview of the models capabilities in section \ref{sec:QualitativeResults}, then benchmark its errors on a test dataset in sections \ref{sec:NoiseDependence} and \ref{sec:MCDropout}. We discuss our results in section \ref{sec:Discussion} and summarize our conclusions in section \ref{sec:Conclusion}.

\section{Methods}\label{sec:Methods}

\begin{figure*}
    \centering
    \includegraphics[width=1\linewidth]{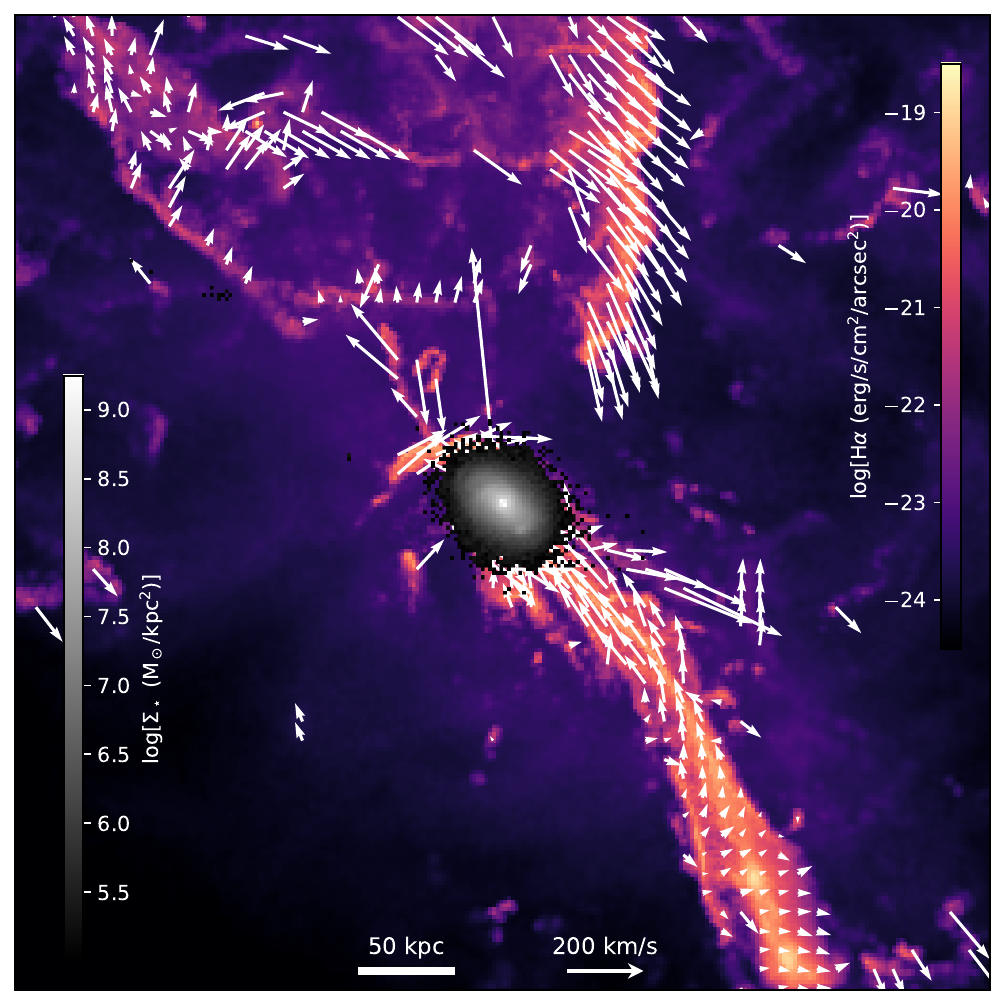}
    \caption{Stellar mass surface density, H$\alpha$ emission, and cold gas velocity of TNG50 subid 392277. $\Sigma_\star$ is plotted on top of H$\alpha$ emission. A quiver plot is overlayed on the H$\alpha$ emission to show the velocities associated with cold gas structures in the CGM.}
    \label{fig:emission maps}
\end{figure*}

\texttt{UNet} is a convolutional encoder--decoder neural network architecture with skip connections that maps images to images. In the original implementation of \texttt{UNet}, the output array is trained to segment an image into distinct regions, with each pixel assigned a probability of belonging to a region \citep{Ronneberger2015_UNet}. For our implementation we instead train the model to match two output arrays to the true plane-of-sky velocity of cold gas in the simulated CGM. We use projections of cold CGM gas from the IllustrisTNG50 \citep{TNG50-1,TNG50-2} cosmological simulation (Sec.~\ref{sec:IllustrisTNG50}) to generate maps of cold gas emission (Sec.~\ref{sec: Emission maps}) and velocity (Sec.~\ref{sec: velocity maps}). The network architecture is described in Sec.~\ref{sec:Network Architecture}, and training and optimization is described in Sec.~\ref{sec: Training and Opt}.

\subsection{IllustrisTNG50}\label{sec:IllustrisTNG50}
The IllustrisTNG project \citep{TNGMethods-2,TNGMethods-1} consists of 3 large-volume cosmological magnetohydrodynamic (MHD) simulations of galaxy formation, generated using the  moving-mesh \texttt{Arepo} code \citep{Springel2010_AREPO,Arepo}. It includes prescriptions for baryonic feedback, with galactic winds dependent on star formation rate \citep{TNGMethods-1}. AGN feedback is split between a high accretion thermal mode (quasar mode) and a low accretion kinetic mode (maintenance mode) expected to drive and distrupt CGM flows at lower redshifts \citep{TNG_physics1}. Neutral Hydrogen fractions are calculated using a spatially uniform UV background accounting for self shielding within each gas cell \citep{Katz1996_H1_atomic_model,Faucher2009_TNG_UVbackground,Rahmati2013_H1_column_density_self_shielding,Nelson2018_TNGresults1_mentions_H1} and gas flows account for the effects of magnetic fields \citep{TNG_magneticFields}. We use the smallest volume ($\sim(50\text{ cMpc})^3$), highest resolution ($\sim8\times10^4M_\odot$ in baryons) simulation: TNG50 \citep{TNG50-1,TNG50-2}.

We exclusively use TNG50 snapshot 99, corresponding to $z = 0$, and select galaxies from the MW/M31-like sample identified in \cite{Pillepich2024_MWsample}. The sample selects for galaxies with stellar masses between $10^{10.5-11.2}\text{M}_\odot$ that are separated from other galaxies of similar or greater mass by at least $500$~kpc, disky, and fall near the center of their host halos. We use this sample because it guarantees that our galaxies of interest will dominate the mass in each of our generated images, and because the CGM of these galaxies have already been analyzed by \cite{Ramesh2023_TNG50CGM1,Ramesh2023_TNG50CGM2}. The full sample includes 198 galaxies. Since they would not contribute to loss calculations, we discard galaxies with no significant CGM H$\alpha$ emission (described in Sec.~\ref{sec: Emission maps}), reducing the sample to 182 galaxies.

The simulated CGMs of TNG50 galaxies have been extensively compared to observations. \cite{Pillepich2021_XrayBubblesFeedback} found that TNG50's feedback reproduces Fermi bubbles observed in X-ray emission; \cite{Byrohl2021_TNG50_MUSE_La_comparison,Nelson2021_MgIIEmissioCGMTNG50} compared stacked emission in Lyman $\alpha$ and MgII; and \cite{Nelson2020_Small_scale_TNG50_Cold_CGM,DeFelippis2021_MgII_absorption_MEGAFLOW_TNG50_compare} found agreement in cold gas covering fractions compared to QSO absorption observations. However, in comparing to unstacked HI emission detected with MeerKAT, \cite{Marasco2025_H1_TNG50_FIRE_MHONGOOSE_compare} found that TNG50 gas was more turbulent and less regular in emission than observations, although this disagreement lessened when comparing more regular, disky galaxies. 

The CGM is a regime in which simulations are still in disagreement \citep{Wright2024_BaryonCycleSimulations,Medlock2025_CAMELSBaryoneCycle}. A full comparison between simulations is beyond the scope of this paper, and we make the assumption that the CGMs in TNG50 are close enough in describing the CGMs of real galaxies to be useful for interpreting H$\alpha$ measurements from upcoming observations. While a cross comparison between simulations will be valuable in the future, it is beyond the scope of the current work, which focuses on introducing and vetting the CNN method. We note that while some simulations have higher spatial resolution or potentially useful comparative physics (e.g., FOGGIE \citep{Peeples2019_FOGGIE}, FIRE-2 \citep{Hopkins2018_FIRE2,Wetzel2023_FIRE2}), TNG50 is ideal for this study as it has the largest number of galaxies in this mass range at reasonable resolution in the CGM, a prerequisite for training neural networks.

\subsection{Dataset Generation}\label{sec:Dataset Generation}

\begin{table}[]
    \centering
    \begin{tabular}{ccc}
        $x$ & $y$ & $z$ \\
        \hline
        0.00 & 0.00 & 1.00 \\
        0.50 & 0.00 & 0.87 \\
        0.87 & 0.00 & 0.50 \\
        1.00 & 0.00 & 0.00 \\
        0.00 & 0.50 & 0.87 \\
        0.00 & 0.87 & 0.50 \\
        0.00 & 1.00 & 0.00 \\
        0.50 & 0.87 & 0.00 \\
        0.87 & 0.50 & 0.00 \\
        0.50 & 0.50 & 0.71 \\
    \end{tabular}
    \caption{Sightlines used to generate 2D emission and velocity maps. $x$, $y$ and $z$ are the fundamental coordinates used by the TNG50 simulation. Throughout the rest of this paper, we refer to coordinate axes $u$, $v$, and $w$, where $w$ is the direction along the line of sight.}
    \label{tab:line-of-sight}
\end{table}

We generate our data using spherical cutouts with 500~kpc radii around galaxies in the MW/M31 sample in the $z=0$ snapshot of TNG50. Each cutout was then imported into \texttt{yt} \citep{YT_methods}, which we used to generate every 2D map and profile used in training. \texttt{yt} visualizes \texttt{Arepo}'s moving mesh data as a set of particles with associated volumes, and smooths values associated with each particle over a length given by
\begin{equation}
    h_{yt}=f_{\textrm{smooth}}\left(\frac{3V}{4\pi}\right)^{1/3}
\end{equation}
where $V$ is the volume associated with each particle, and $f_{smooth}$ is a smoothing factor. We use \texttt{yt}'s default $f_{smooth}=2$, but note that the exact value of $f_{smooth}$ does not affect our results due to our analysis focusing on cold, dense gas, and \texttt{TNG50} gas particles having similar masses regardless of density, the particles that contribute to our emission and velocity maps are smoothed to resolutions finer than our $~2$kpc/pix map resolution.

For each galaxy in our sample, we generate emission and velocity maps along 10 sight lines, each from the 1st xyz octant. The normalized sight line vectors for each map are shown in Table~\ref{tab:line-of-sight}. For each sight line map, we rotate \ang{0}, \ang{90}, \ang{180}, and \ang{270} and optionally reflect over the vertical axis for a total of 8 views per line of sight, or 80 views per galaxy. While the 1D mass and flow profiles are independent of viewing angle, the 2D velocity maps change for each line of sight, so our total dataset consists of $80\times182=14560$ distinct (although still correlated) input/output pairs.

\subsubsection{Emission Maps} \label{sec: Emission maps}
We generate emission maps of H$\alpha$ following the procedure of \cite{Nelson2021_MgIIEmissioCGMTNG50} (their Sec.~2), which we briefly describe here. Line volume emissivities of H$\alpha$ were computed using \texttt{Cloudy} \citep{Cloudy} over values in a grid of $n_H,T,Z$ (hydrogen number density, temperature, and metallicity). The emissivity of each particle in the simulation is set by interpolating between values on the \texttt{Cloudy} grid. We then multiply by the gas species fraction relative to solar to obtain an emission density [erg/s/cm$^3$], and divide by $4\pi$ steradian to get a brightness in [erg/s/cm$^3$/arcsec$^2$]. We integrate emission along the line-of-sight on to a $256\times256$ pixel grid, with side lengths of 500kpc to get a final map of emission in [erg/s/cm$^2$/arcsec$^2$]. By integrating in this way we are making the assumption that the CGM is optically thin along every sight-line, and that the clumping factor $C=\langle n^2\rangle/\langle n\rangle^2=1$ within each cell. Example emission maps in H$\alpha$ for a test galaxy are shown in Fig.~\ref{fig:emission maps}.

Our emission maps are modified by noise to simulate what real data might look like. \mb{We generate an array of Gaussian noise over the entire $256\times256$ pixel grid which is then added to the emission maps.} In all cases, even when noise is not added, emission is clipped to a lower bound of $10^{-25}$erg/s/cm$^2$/arcsec$^2$. For training and validation (discussed further in Sec.~\ref{sec: Training and Opt}) we use two versions of every galaxy view: one with no noise, and one with a random amount of noise between $-22<\text{log}(\text{noise})<-19.5$ taken from a uniform distribution in log space. To regularize the data, we take log(emission/erg/s/cm$^2$/arcsec$^2$)$+23$ as the emission inputs to our model.


\subsubsection{Cold Gas Velocity Maps}\label{sec: velocity maps}
We calculate a 2D map of the plane-of-sky cold gas ($T<10^{4.5}K$) velocity, averaged along the line of sight of each pixel on the same $256\times256$ pixel grid we used for emission maps. 


For a line of sight $\hat{\boldsymbol{w}}$ we calculate $v$ in directions $i\in[u,v]$ using
\begin{equation}
    v_i(\textbf{X}) = \int_{T<10^{4.5}} v_i(\boldsymbol{x})\tilde{F}_{H_\alpha}(\boldsymbol{x})\hat{\boldsymbol{w}}\cdot d\boldsymbol{x}
\end{equation}
where $\tilde{F}_{cold}(\boldsymbol{x})$ is a weight given by
\begin{equation}
    \tilde{F}_{H_\alpha}(\boldsymbol{x})=\frac{F_{H_\alpha}(\boldsymbol{x})}{\int_{T<10^{4.5}} F_{H_\alpha}(\boldsymbol{x})\hat{\boldsymbol{w}}\cdot d\boldsymbol{x}}.
\end{equation}

We weight based on H$\alpha$ emission so that a particular gas cell will be equally represented in both the input and output of the model. We note that this method of velocity projection fails if there is no cold ($T<10^{4.5}K$) gas along a particular line of sight. In these cases we internally record the velocity as 0 in both directions, but note that due to the loss mask used in training (Sec.~\ref{sec: Training and Opt}) errors along these lines of sight are never evaluated and do not affect the model.

\subsection{Network Architecture}\label{sec:Network Architecture}
Our CNN uses a UNet architecture \citep{Ronneberger2015_UNet}, implemented in \texttt{PyTorch} \citep{pytorch} by \texttt{Segmentation Models} \citep{segmodels}. To summarize, a UNet consists of an encoder and a decoder. The encoder consists of a series of convolutional layers, each reducing the size of each channel while increasing the number of channels, so at the lowest layer of the encoder, the data is the most encoded with the lowest level of spatial resolution. The decoder works in reverse: each convolutional layer increases the size of each channel while reducing the number of channels. Additionally, the output of each layer of the encoder is appended to the input of each layer of the decoder of the same depth. A final convolutional head reduces the number of output channels to our two velocity components. 

\subsection{Training and Optimization}\label{sec: Training and Opt}
We divide our 182 galaxies into a 70\%/15\%/15\% train/test/validation split, resulting in 126 training galaxies, 28 test galaxies, and 28 validation galaxies. Training galaxies are used train model weights, validation galaxies are used to determine the best model checkpoint and tune hyperparameters, and the performance of the model on test galaxies is discussed and benchmarked in Sec.~\ref{sec:Results}. The different views, rotations and reflections of a given galaxy are all correlated, but by splitting our data at the galaxy level we guarantee that there is no correlation between the the train, test, and validation sets. No data used in the validation set comes from galaxies in the training set, and no data used in the testing set comes from galaxies used to train or validate models.

For the purpose of calculating loss, we mask a large portion of the field of view in each sight line. Pixels with little to no cold gas emission have disordered or undefined cold gas velocities, so we make a \mb{surface brightness} cut using bolometric H$\alpha$ emission, requiring $L_{\textrm{H}\alpha}>10^{-21}$erg/s/cm$^2$/arcsec$^2$. Additionally, cold gas within the host galaxy has a much greater typical velocity than gas in the CGM, with structure much smaller than what can be resolved by our (256,256) pixel grid, \mb{so we also mask a circle with a radius of 8 pixels} ($\approx16$kpc) of the center of the galaxy for each sight line. The combined mask determines which pixels in each sight line will be used for loss calculation when model training. \mb{We use the same mask when characterizing model errors (Sec. \ref{sec:NoiseDependence}), but note that we never mask model inputs, so the model can see gas emission in the central 8 pixels even though its predictions for those pixels are ignored.} For the remaining H$\alpha$-bright, non-central pixels, the loss is given by the MSE (mean squared error):
\begin{equation}\label{eq:loss}
\begin{split}
    \mathcal{L}_{los} = \langle(v_{u,\textrm{true}}-v_{u,\textrm{pred}})^2+(v_{v,\textrm{true}}-v_{v,\textrm{pred}})^2\rangle_{\textrm{valid pixels}}
\end{split}
\end{equation}
where $u$ and $v$ are orthogonal components of the plane-of-sky velocity, pointing along the axis of our pixel grid described in Sec.~\ref{sec:Dataset Generation}.

We also experimented with instead determining loss by first converting the $u,v$ components of the plane-of-sky velocities into an angle $\theta$ and a magnitude, and then calculating loss as 
\begin{align}
    \mathcal{L}
    =
    &\lambda_\theta(-\cos(\theta_{\textrm{true}}-\theta_{\textrm{pred}}))+
    \notag\\&\lambda_{\textrm{mag}}(\textrm{mag}(u,v)_{\textrm{true}}-\textrm{mag}(u,v)_{\textrm{pred}})^2 
\end{align}
However, initial testing suggested this approach worked worse than our initial $u$,$v$ approach and parameterizing loss in this way would require additional (and somewhat arbitrary) balancing of the $\lambda_\theta$,$\lambda_{\textrm{mag}}$ parameters since the angle and magnitude components of the loss are on different scales.

As initial weights, we use the \texttt{ImageNet} \citep{ImageNet} pretrained encoder weights implemented in \texttt{Segmentation Models}. Using pre-trained encoder weights helps models converge faster to more stable weights. The initial decoder weights are random. We freeze encoder weights for the first 3 epochs of training so that the decoder can settle, then unfreeze the encoder to train the entire model.

During training, we minimize the loss function (Eqn.~\ref{eq:loss}) of galaxies in our training set over a number of epochs. We train using the \texttt{PyTorch} implementation of \texttt{ADAM} \citep{kingma2014adam}, calculating the validation loss after each epoch and take our best fit model realization to be the one that minimizes validation loss. We treat \texttt{ADAM}'s weight decay and initial learning rate, as well as the decoder channel width, dropout fraction, and pretrained encoder as hyperparameters, which we optimized with \texttt{Optuna} \citep{optuna}. Hyperparameter ranges and best fit values are presented in Appendix~\ref{sec:hyperparameter optimization}. In practice, most training runs reach a minimum validation loss after $\sim20-30$ epochs, so we let each each \texttt{Optuna} trial run for 30 epochs, and do our final training with the best fit hyperparameters over 50 epochs.

\mb{We note here that much about the model's training and optimization could be optimized or tweaked. For instance, stellar surface brightness could be used to mask the galaxy's ISM rather than the somewhat arbitrary circular mask used, or the weights of each pixel could be adjusted based on gas mass, or so that the total weight given to outflows equals the weight given to inflows. These adjustments are worth investigating in future papers, but we justify our approach as one that does not use any information in training other than H$\alpha$ emission maps, so that any capabilities of the model can be attributed only to information contained in those maps.}

\section{Results}\label{sec:Results}
\subsection{Qualitative Capabilities}\label{sec:QualitativeResults}

\begin{figure*}
    \centering
    \includegraphics[width=0.95\linewidth]{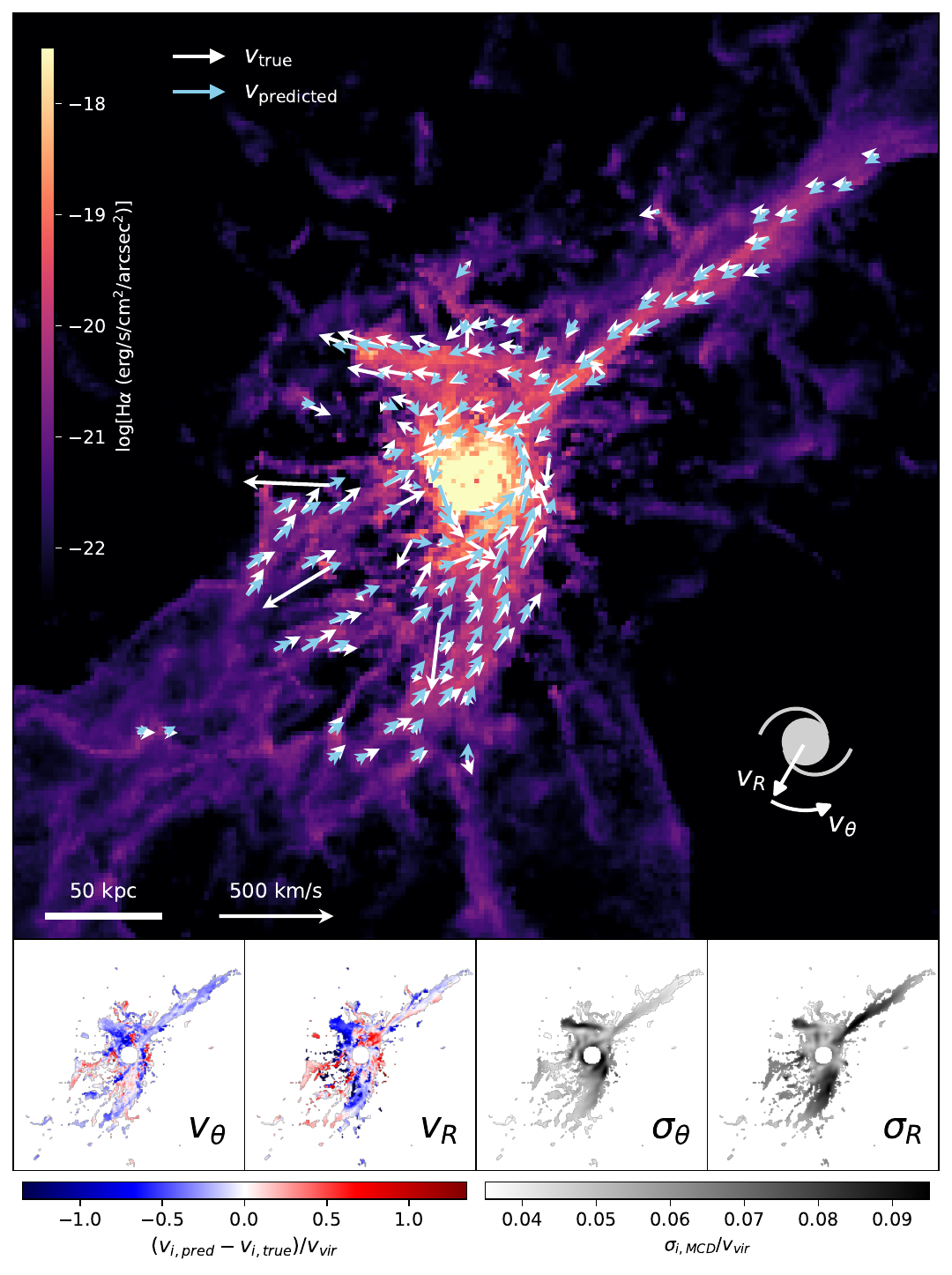}
    \caption{Top: Quiver plot of predicted (blue) and true (white) values for the plane-of-sky velocity of cold gas around a galaxy in our test set (TNG50 subid 552581), overlayed on its H$\alpha$ emission. Bottom Left: Signed errors in the predicted $v_\theta$ and $v_R$ over the regions of the CGM that would have been used to determine loss if the galaxy were in our training set. Bottom Right: Standard deviations of Monte Carlo Dropout draws, highlighting regions of the predictions with high epistemic uncertainty.}
    \label{fig:552581 basic}
\end{figure*}

\begin{figure*}
    \centering
    \includegraphics[width=1\linewidth]{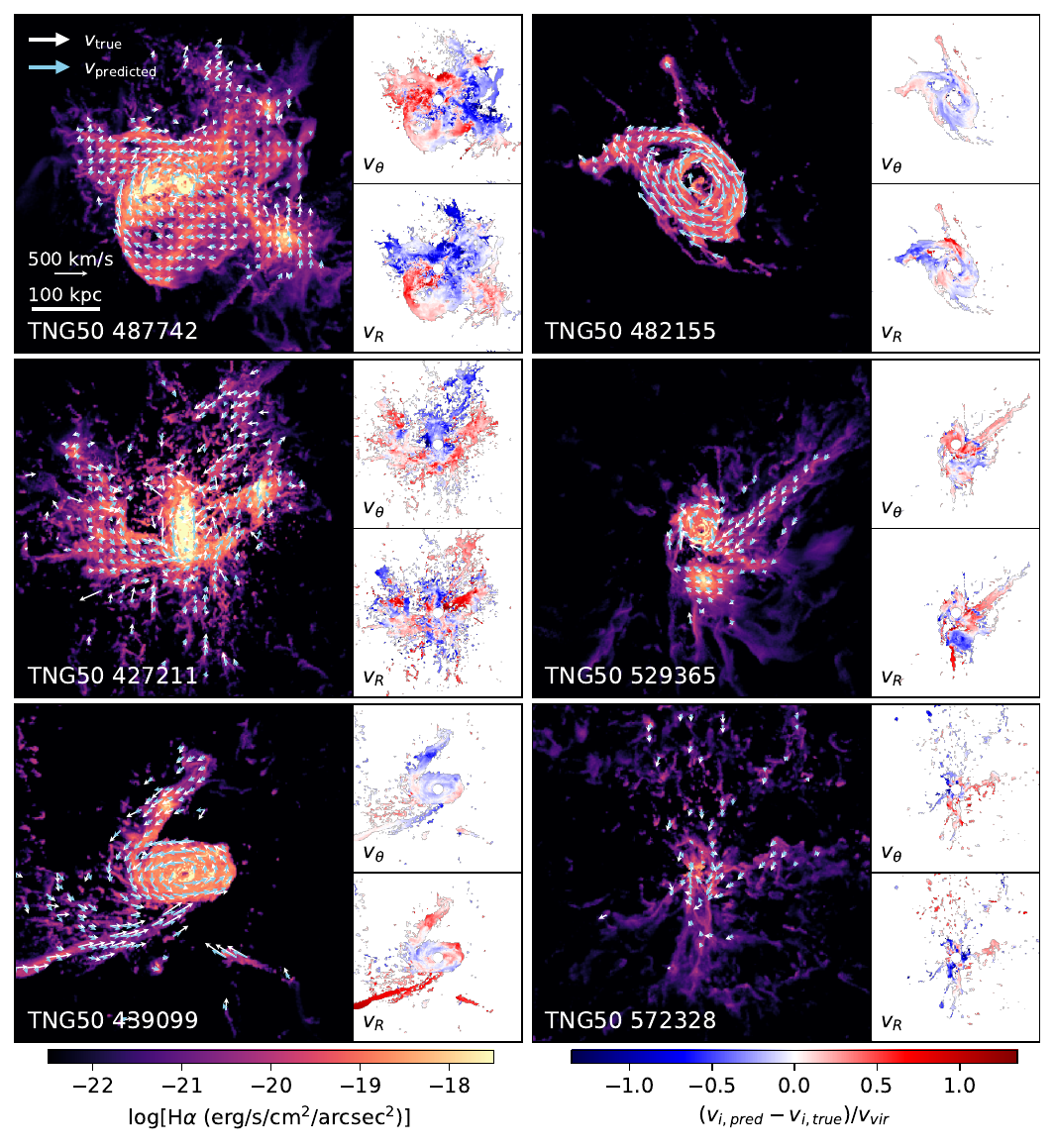}
    \caption{Quiver plots of predicted (blue) and true (white) values for the plane-of-sky velocity of cold gas around six galaxies in our test set, chosen to showcase a range of CGM morphologies. TNG50 subids are shown for each. Subids 487742 and 427211 show the response of the model to ``messy" CGMs with multiple satellites. For these, the model has moderate predictive power for the broader, more cohesive flows but smooths over small deviations and generally has larger errors than for CGMs with simpler morphologies. 439099, 482155, and 529365 have simpler morphologies with distinct disks and inflows; the model performs better in these cases, always identifying the disk rotation, velocity gradients in inflows, and correctly inferring the flow of gas on the left side of 482155 that doesn't flow along the line formed by the gas. 572328 has a more limited "spotty" morphology. Without as much visible structure as the other examples the model has a harder time inferring velocities, especially near the center.}
    \label{fig:test sample quiver}
\end{figure*}

The results of our model applied to a sample view of a galaxy in our test set is shown in Fig.~\ref{fig:552581 basic}. This particular galaxy and view were chosen because its shows a handful of distinct CGM features that can be compared to illustrate where the model works, and where it does not. In the top right of the view is a narrow inflow pointing almost directly towards the host galaxy. The bottom left has a wider, patchier inflow, which has more tangential velocity as it approaches the galaxy. Gas associated with a small satellite can be seen above the central galaxy, and cold gas associated with polar outflows can be identified in the bottom left by its large radial velocities. The model is able to identify the inflows, including the tangential component of the bottom inflow, and distinguishes the gas associated with the satellite from the inflows, although it does a poorer job of estimating its velocity. It is able to infer the velocity gradient in the top right inflow, with lower velocities at higher radii. It also correctly identifies the overall rotation of gas close to the central galaxy. However, it is unable to separate the gas assosiated with the polar outflow from the bottom left inflow, and in general smooths over the small scale velocity variations in the bottom inflow. It estimates tangential velocities to a typical error of $\sim 0.3~v_{\textrm{vir}}$, with similar errors in radial velocity along the inflows that jump to much larger errors for the satellite and especially the polar outflow. 

A sample of six more test galaxy views are shown in Fig. \ref{fig:test sample quiver}. The qualitative pattern appears to be that the model correctly identifies the sign of the overall rotation of gas around the central galaxy, if there is one. It usually can infer the velocities of inflows, including those with significant tangential components. It is sometimes confused by flows that do not move in the same direction as lines formed by the gas (e.g., the top clump in subid 487742), but sometimes correctly infers the direction of such flows (e.g., the left of subid 482155, the middle right of subid 529364). It does a poorer job with gas associated with satellites, and the presence of multiple satellites may lead to larger errors across the entire image (e.g., subids 487742 and 427211). It never infers the velocities of gas associated with polar outflows to any accuracy (some are visible in subid 427211), and overall tends to ``smooth over" small scale velocity variations, even if they are not associated with clear outflows (e.g. subid 427211). 



\subsection{Noise Dependence}\label{sec:NoiseDependence}
\begin{figure*}
    \centering
    \includegraphics[width=1\linewidth]{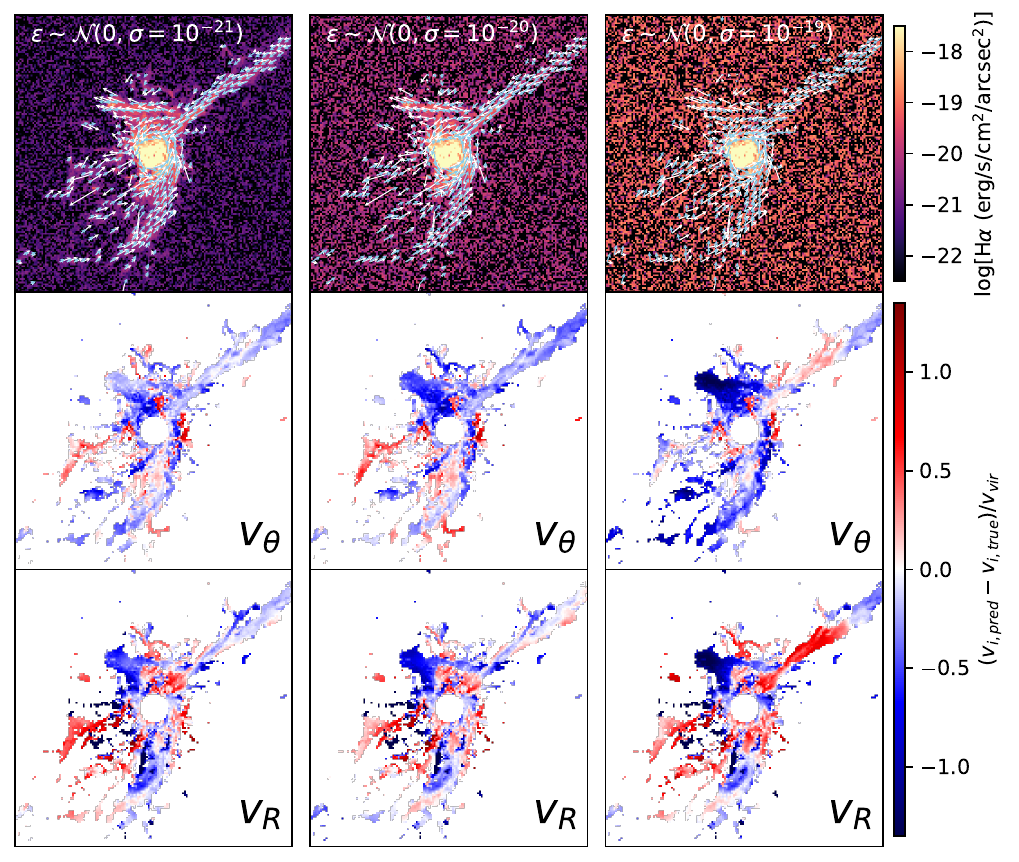}
    \caption{Dependence of model predictions on increasing Gaussian noise. Each column shows the results for a specific noise level. Top: input H$\alpha$ map, true velocities (white) and predicted velocities (blue). Middle and Bottom: signed errors in $v_\theta$ and $v_R$. For noise $\epsilon\sim N(0,\sigma=10^{-21})$ much of the CGM structure is still visible and the errors do not increase noticeably compared to the zero noise case (see Fig.~\ref{fig:552581 basic}. For noise $\epsilon\sim N(0,\sigma=10^{-20})$ errors in the satellite and patchier inflows increase, but the broader features maintain similar errors to the low noise case. By $\epsilon\sim N(0,\sigma=10^{-19})$ most features in the CGM are not detectable and the model defaults to a prediction of global radial infall.}
    \label{fig:single_galaxy_noise}
\end{figure*}

\begin{figure*}
    \centering
    \includegraphics[width=1\linewidth]{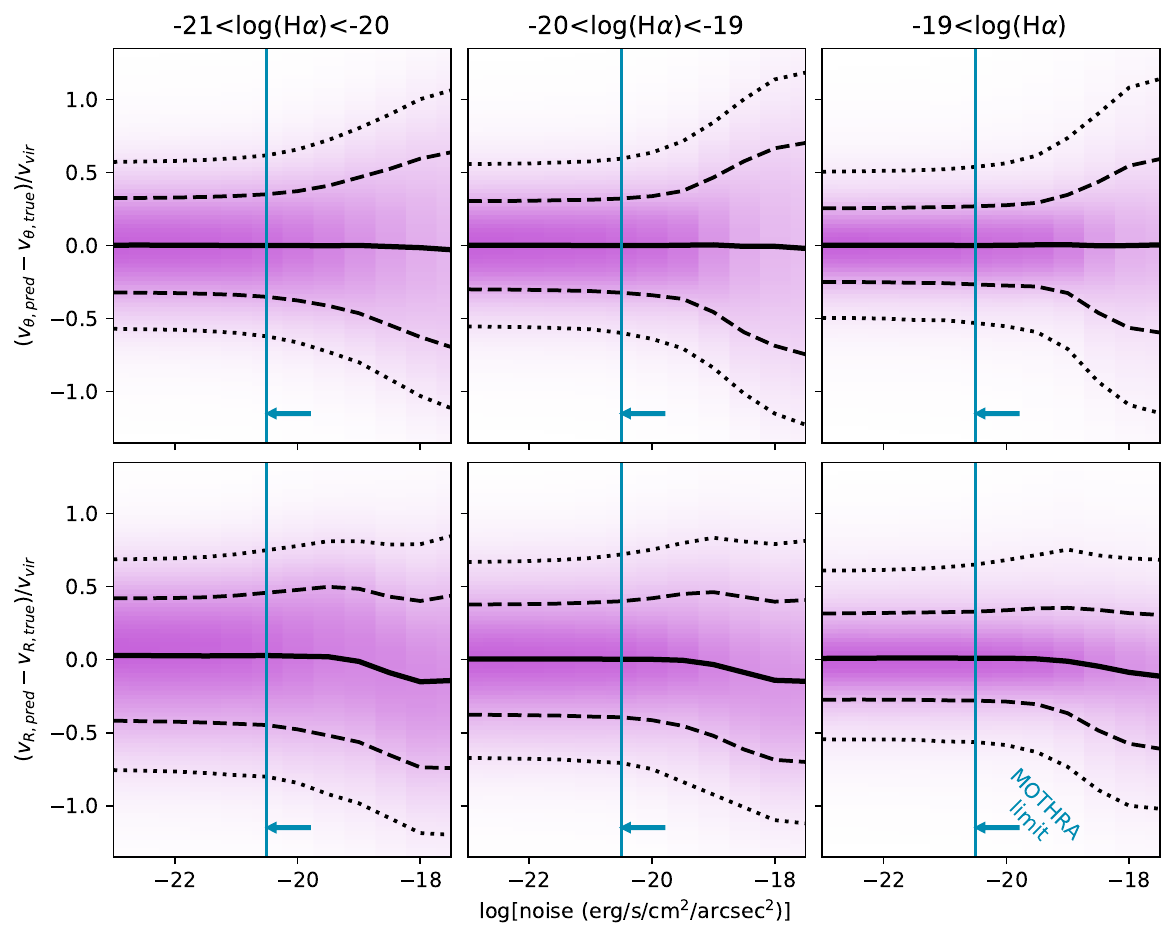}
    \caption{Dependence of the signed error in $v_\theta$ and $v_R$ on Gaussian noise. Error distributions were calculated using the set of all H$\alpha$-bright, non-central pixels that satisfy the loss mask (Sec.~\ref{sec: Training and Opt}) in the set of test galaxies. Lines trace the 5th, 16th, 50th, 84th and 95th percentiles. Pixels are binned by their zero noise H$\alpha$ emission, brighter pixels are better predicted at low noise but as noise increases pixels of all brightnesses increase in error at around the same noise level of $\sigma\sim10^{-20}$. The anticipated MOTHRA detection limit is shown in each plot.}
    \label{fig:noise_diagnostic}
\end{figure*}

The effects of increasing Gaussian noise on a single test galaxy are shown in Fig.~\ref{fig:single_galaxy_noise}, and summary distributions of the entire test set are shown in Fig.~\ref{fig:noise_diagnostic}. There is a clear inflection point around a noise level of $10^{-20}$~erg/s/cm$^2$/arcsec$^2$/\AA, below which the median absolute $v_\theta$ error is $\sim40$~km/s, and above which it doubles to a value of $\sim80$~km/s, with a smaller increase in the absolute $v_R$ error at the same noise level. This inflection point occurs at the same noise level regardless of the noise-less emission of the pixel in our test set: the velocities of pixel with emission $<10^{-20}$ are still predicted moderately well when their individual signal to noise ratios are less than 1, and pixels with emission $<10^{-19}$ still see an increase in error around a noise level of $10^{-20}$, despite still being easily detectable.

The cause of the inflection in Fig.~\ref{fig:noise_diagnostic} can be seen by observing the changes noise has on a single galaxy in Fig.~\ref{fig:single_galaxy_noise}. As noise increases, more and more of the broader CGM structures are not detectable. While the largest inflows are still visible, the network is able to leverage patterns at larger scales to make predictions about gas below the detection limit. For the face-on view of subid 552581, the two large inflows as well as the gas associated with the satellite galaxy are still visible at a noise of $10^{-20}$, even if the finer details in the structure are not, so the network is still able to infer the curl of the gas as it approaches the galaxy. However, at a noise of $10^{-19}$ only the gas associated with the ISM of the galaxy and the satellite are visible and the network assumes that all cold gas is radially infalling towards the central galaxy, including the small clump associated with the satellite that is still visible. The critical noise level seen in Fig.~\ref{fig:noise_diagnostic} is just the characteristic emission of the CGM gas in our sample. We note that the predicted depth of the MOTHRA telescope is comfortably below that threshold.

The high noise case can also be interpreted as a prior in the model: this is what the model predicts when it can detect a galaxy's ISM but not its CGM. The difference between this prior and the low noise case could be interpreted as what the model has actually learned from the morphology of the CGM. With this in mind we can compare to the qualitative results discussed in Sec.~\ref{sec:QualitativeResults}. The model is quite good at predicting the curl of the gas around the central galaxy, which corresponds to the large decrease in $v_\theta$ errors, but results are more mixed when it comes to predicting feedback driven outflows, which explains the more moderate decrease in the $v_R$ errors.

\subsection{Bayesian Dropout}\label{sec:MCDropout}


\begin{figure}
    \centering
    \includegraphics[width=1\linewidth]{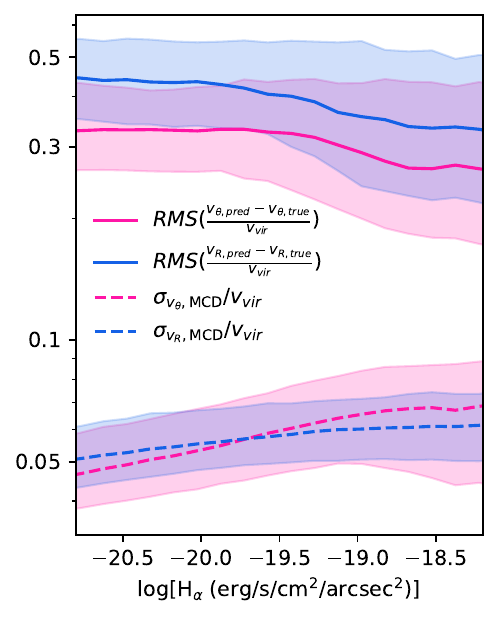}
    \caption{Dependance of root mean squared (RMS) errors and Monte Carlo dropout (MCD) deviation on pixel brightness for the zero noise test set of galaxy views. Error widths are the 16th and 84th percentiles of the RMS and MCD deviation over the test sample. The RMS errors are significantly larger than MCD deviation at all levels, indicating that \mb{aleatoric} uncertainties are the dominant source of error in our model. RMS errors show a moderate decrease with increasing brightness, likely due to bright pixels having more visible substructure around them for inference. MCD deviation increases with brightness due to brighter pixels being less common in the training dataset and therefore having larger assosiated epistemic uncertainties.}
    \label{fig:MC_dropout_diagnostic}
\end{figure}

We further characterize the network using Monte Carlo Dropout (MCD), a method presented in \cite{Gal2015_MCDropout}. The core idea of MCD is that, by taking different realizations of the dropout layers in the decoder of the model we can get a distribution of outputs for each input that cover a space of equally likely realizations of the model given the training set. The variation in this output space is a measure of the model's epistemic uncertainty: a systematic uncertainty associated with the model itself due to the combination of a limited training dataset and limitations of the model architecture. Epistemic uncertainty can be contrasted with aleatoric uncertainty at the data level, which cannot be reduced with more data or better model selection \citep{DeeplyUncertain}.

Fig.~\ref{fig:MC_dropout_diagnostic} compares the behavior of the RMS velocity errors and deviation due to MCD on the emissivity of the pixel being predicted. The RMS error is consistently much larger, which is as expected: by viewing the CGM from a single direction using only H$\alpha$ emission we eliminate any chance of there being a one to one mapping of inputs to outputs, even before we consider the effect of noise. This will manifest as a large aleatoric, systematic uncertainty, so for our model the standard deviation of MCD draws ($\sigma_{i,\textrm{MCD}}$) has no significant corelation with the actual errors. Instead, $\sigma_{i,\textrm{MCD}}$ is seen to increase with the H$\alpha$ emission of the measured pixel. This might be explained by brighter pixels being less common in the training set: the model has seen less features, and so has more disagreement with itself on how to interpret them. Conversely, bright pixels tend to be surrounded by dimmer pixels with considerable substructure that the network can leverage to make a more informed inference so the actual RMS errors of pixels tend to decrease with brightness. This behavior can also be seen in lower right subplots of Fig.~\ref{fig:552581 basic}. \mb{The model's self-disagreement,} $\sigma_{\theta,\textrm{MCD}}$, increases around the regions of the inflows that are not radially infalling, suggesting that the morphologies of these features are strongly informing the inferred $v_\theta$ and that specific realization of these morphologies are less common in the training set. The narrow inflow and patchier inflow have greater $\sigma_{R,\textrm{MCD}}$ for the same reasons, and the satellite has greater $\sigma_{\theta,\textrm{MCD}}$ and $\sigma_{R,MCD}$ since its morphology is both less common, pixel wise, than inflows and it strongly affect both its $v_\theta$ and $v_R$.

We note that MCD is not not a perfect estimator of epistemic uncertainty. \cite{Behavior_of_MCdropout} found that the posterior estimated by MCD depends significantly on the dropout fraction in dropout layers. Ensembling \citep{Ensembling} has been proposed as an alternative method, and \cite{MCdropout_ensambling_compare} found that it outperforms MCD over a variety of test cases. However, given our code, structure ensembling would require significantly ($\sim 100\times$) more computation time than MCD. Since we expect aleatoric uncertainties to dominate errors, we elect to use MCD in this analysis.

\section{Discussion}\label{sec:Discussion}
\subsection{Use Case and Limitations}

The model presented in this paper is intended primarily as a proof-of-concept, and demonstrates that (i) the 2D emission morphology of cold CGM gas has a mapping to its plane-of-sky kinematics (although not a one-to-one mapping); (ii) a CNN can learn this mapping; (iii) there is a critical noise level above which this mapping fails; which corresponds with the standard emission level of the CGM; and (iv) the signal-to-noise needed to take advantage of this relationship is achievable by upcoming telescopes, all under the assumption that actual CGM gas looks and behaves similarly to what is found seen in TNG50. We note the caveat that the relationship between 2D morphology and kinematics is clearly not one-to-one, and interpretation of a model output 2D velocity map may be difficult if one expects, for instance, two disconnected gas clouds with different velocities occupying the same line-of-sight. As is, using this model in practice would require careful consideration of the data to decide whether the model's output regarding a particular clump of gas is reasonable.


While there are numerous possible ways to improve the methods discussed in this paper, the largest caveat is that, as is the case with any simulation-based inference, any confidence in the applicability of the model to real world data assumes that the simulations used accurately reflect the real world. While TNG50 has decent agreement with current limits from CGM observation \citep{Nelson2020_Small_scale_TNG50_Cold_CGM,Byrohl2021_TNG50_MUSE_La_comparison,Pillepich2021_XrayBubblesFeedback,DeFelippis2021_MgII_absorption_MEGAFLOW_TNG50_compare}, the large radius, smaller scale morphology that our model uses cannot be compared because it has not been observed yet, and this structure varies significantly between simulations \citep{Wright2024_BaryonCycleSimulations,Medlock2025_CAMELSBaryoneCycle}. Additionally, we do not attempt to model subgrid effects when calculating emission e.g. if the unresolved clumping factor $C=\langle n^2\rangle/\langle n\rangle^2$ is $>>1$ or significantly non-uniform, then our simulated emission will be systematically dimmer and/or more uniform than true observations. While comparing the results of models applied to different simulations than what they were trained on is worthwhile, a direct comparison between observed H$\alpha$ morphology and TNG50 and similar simulations will be necessary before any simulation based inference can be used.

\subsection{Future Work}
There are likely many ways to improve our model, decisions made about the specific of our inputs, outputs, and loss calculations were made with the goal of isolating emission morphology as the sole input with no attempt to classify/distinguish/weight regions of the CGM. For instance, we make no attempt to balance inflows and outflows in our loss calculations, instead opting to weight all pixels equally. It is standard when training classifiers to include equal amounts of each class in training and validation sets, so as a classifier of inflows vs outflows, our model is non-optimal. Morphologically classifying structures in the CGM, and weighting loss based on membership in those groups rather than on a per-pixel basis, might help inference for less common structures. Training a model to identify properties of individual clouds could also permit training on higher resolution, zoomed in simulations \citep[e.g.,][]{Gronke2020_cloudcrushing,Ramesh2026_zoomincloudsim} which might reproduce the cloud morphology better than cosmological simulations.

The currently used training sample of MW/M31-like galaxies is a good starting point, but to apply the model to a wider range of real observations, the training sample needs to span a wider range of galaxy masses, star formation rates, and environments. The lower sample size of massive quiescent galaxies in cosmological simulations complicates this, but the training sample might be increased by using snapshots $\Delta t>t_{\textrm{dynamic}}$ in the past of the same galaxies used in the $z=0$ snapshot. A larger training set will likely help inference for less common morphological features as identified in Sec.~\ref{sec:MCDropout}.

We identified in Sec.~\ref{sec:MCDropout} that epistemic uncertainty does not dominate our errors, and in Sec.~\ref{sec:NoiseDependence} we identify that aleatoric statistical errors due to noise only begin to dominate at observing depths $>10^{-20}$ erg/s/cm$^2$/arcsec$^2$. MOTHRA can probe deeper than this critical depth, so we expect the largest gains in accuracy could be achieved by minimizing aleatoric systematic errors. \mb{Throughout this paper we intentionally limit ourselves to only utilize H$\alpha$ surface brightness as a model input in order to characterize what information can be gained from H$\alpha$ morphology, but aleatoric errors could be reduced by including complimentary observables as additional model inputs.} For instance, our input contains no estimate for the dynamical mass of the central galaxy; adding this value as an input might significantly help the model. \mb{High resolution line-of-sight velocity information, such as a full data cube, could provide many kinematic constraints.} Adding maps of emission in NII and OIII \mb{(which the upcoming MOTHRA telescope will observe to the same depth as its H$\alpha$ observations)} could add metalicity information, X-ray emission as an input might help constrain outflows of hot gas, \mb{and stellar surface brightness maps could constrain the galaxies inclination and minor axis direction, with strong correlations to the morphology of outflowing gas (e.g. \cite{Guo2023_MgIIOutflow})}. There is also a disconnect between the target velocities (which we define relative to the central galaxy) and the local velocity of the volume filling hot CGM gas that cold gas clouds are actually moving through, which could cause even more issues when trying to predict flows in more complex group environments than those probed by the TNG50 MW/M31 sample. We also note that CNNs are now being outperformed by Vision Transformers (ViT) for large datasets \citep{dosovitskiy2021an_vision_transformer,Mauricio2023_transformer_vs_CNN}, so a change in architecture should be considered.

\section{Summary and Conclusion}\label{sec:Conclusion}
We train a CNN with a \texttt{UNet} architecture to infer plane-of-sky velocities of cold CGM gas using only maps of H$\alpha$ emission as input. The models were trained and tested on the Illustris TNG50 MW/M31-like galaxy simulations, with H$\alpha$ emission calculated using a \texttt{Cloudy} model grid and interpolated and integrated onto a 256 by 256 pixel grid using \texttt{yt}. 80 views of each galaxy were used to sample over many rotations and lines-of-sight. The findings from our optimized model are summarized as follows:
\begin{itemize}
    \item Qualitatively, the model shows the highest accuracy when inferring the velocities of large, regular inflows, with decreased accuracy for features that occupy fewer pixels, such as gas associated with satellite galaxies. It can infer both tangential and radial velocities, and is sometimes able to infer velocities for clouds that are not oriented along their velocity vector. It was able to identify the rotation of gas around the central galaxy in all galaxies in our test sample, but could not identify gas associated with major outflows, likely because this gas is typically hot and therefore invisible to the model.
    \item In the limit of no noise, the model can infer plane-of-sky velocities with RMS errors of $0.3-0.5v_{vir}$.
    \item Errors tend to increase with greater noise and lower pixel emission. There is a critical noise level above which enough CGM substructure cannot be detected where the model defaults to predicting all gas as radially infalling. We interpret this high noise output as the model's learned prior on cold CGM flows, and the increase in accuracy from the high to low noise regime as the information gained from observing the CGM morphology. We note that the MOTHRA telescope is anticipated to observe H$\alpha$ at the requisite depth for inference in the ``low noise" regime.
    \item Utilizing dropout layers in the model decoder, we estimate epistemic uncertainties using Monte Carlo Dropout. The standard deviation in dropout draws is low compared to actual error in model outputs even in the zero noise case, which we interpret as the errors being dominated by aleatoric systematic uncertainties. MCD standard deviations vary over the output, which we interpret as identifying morphological features that are less common in the training set where inference might be improved with a larger training set.
\end{itemize}
These results demonstrate that there is a mapping between observable CGM morphology and not directly observable plane-of-sky velocities, albeit one that is non-linear and not one-to-one. This mapping is learnable by a CNN, and the observations needed to utilize it may soon be obtainable with the MOTHRA telescope. While this has many caveats, the greatest of which is that cosmological simulations might not reproduce the morphology of real CGMs for this type of simulation-based inference to be applicable. If these can be overcome, we can unlock extra dimensions of CGM analysis.

\begin{acknowledgments}

We thank Daisuke Nagai for advice on the use of TNG50, Naomi Gluck and Selim Kalici for instructive conversations and sample code for \texttt{PyTorch}, and Harrison Souchereau and Isabel Medlock for their help with visualizations in \texttt{yt}. C.J. gratefully acknowledges support from a Gruber
Science Fellowship at Yale University. We thank Atlas and Clover for contributing their likenesses for Fig.~\ref{fig:OOD3}.

\end{acknowledgments}



%

\software{\texttt{Cloudy} \citep{Cloudy},
          \texttt{numpy} \citep{numpy},
          \texttt{matplotlib} \citep{matplotlib}
          \texttt{PyTorch} \citep{pytorch},
          \texttt{Segmentation Models} \citep{segmodels},
          \texttt{yt} \citep{YT_methods}, 
          }


\appendix

\section{Hyperparameter Optimization}\label{sec:hyperparameter optimization}

\begin{table}[ht]
\centering
\caption{Hyperparameters explored in the \texttt{Optuna} optimization.}
\label{tab:optuna_hyperparams}
\begin{tabular}{l|c|c}
\hline
\hline
\textbf{Hyperparameter} & \textbf{Search Range} & \textbf{Best Fit} \\
\hline
Encoder architecture &
\{ResNet18, 34, 50, 101, 152\} &
ResNet152 \\

Decoder base width ($D$) &
$\{4,\,8,\,16,\,24,\,32,\,48,\,64,\,96,\,128\}$ &
128 \\

Dropout fraction ($p$) & $[0.4,\,0.05]$ & 0.128 \\

Learning rate ($\alpha$) &
$[10^{-5},\,10^{-3}]$ &
$3\times10^{-4}$ \\

Weight decay ($\lambda_{\mathrm{wd}}$) &
$[10^{-6},\,3\times10^{-4}]$ &
$1.5\times10^{-6}$ \\

\hline
\end{tabular}
\end{table}

Hyperparameters were determined using 100 \texttt{Optuna} trials. Hyperparameter ranges and best fit values are shown in tab. \ref{tab:optuna_hyperparams}. The best fit encoder architecture and decoder size mean that the model used is the largest possible allowable by the hyperparameter ranges, suggesting the true best model might be even larger, which is somewhat concerning. While a full comparison of using different model inputs is outside the scope of this paper, we note that initial tests using NII and OIII emission maps alongside H$\alpha$ decreased the best fit encoder and decoder sizes to fall within the tested range.


\section{Out of Distribution Tests}\label{sec: OOD}
\begin{figure}
    \centering
    \includegraphics[width=0.9\linewidth]{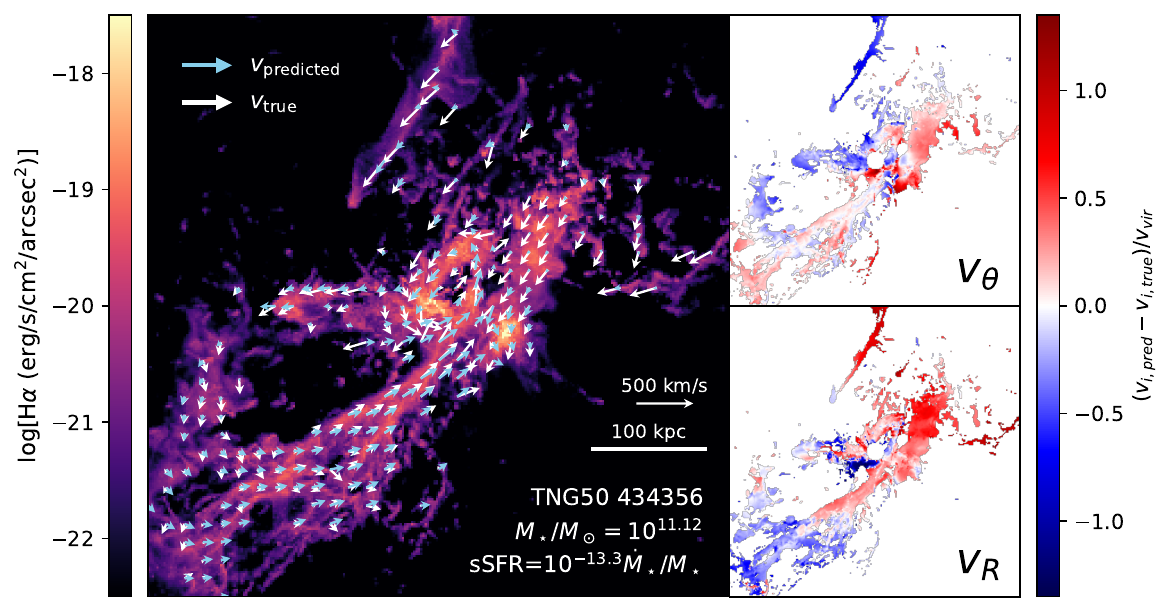}
    \caption{Results of the model applied to a TNG50 galaxy outside of the MW/M31 sample. This galaxy is a massive quiescent, and was selected as an example due to having a comparable CGM morphology to the galaxy shown in Fig.~\ref{fig:552581 basic}.}
    \label{fig:OOD}
\end{figure}
\begin{figure*}
    \centering
    \includegraphics[width=1\linewidth]{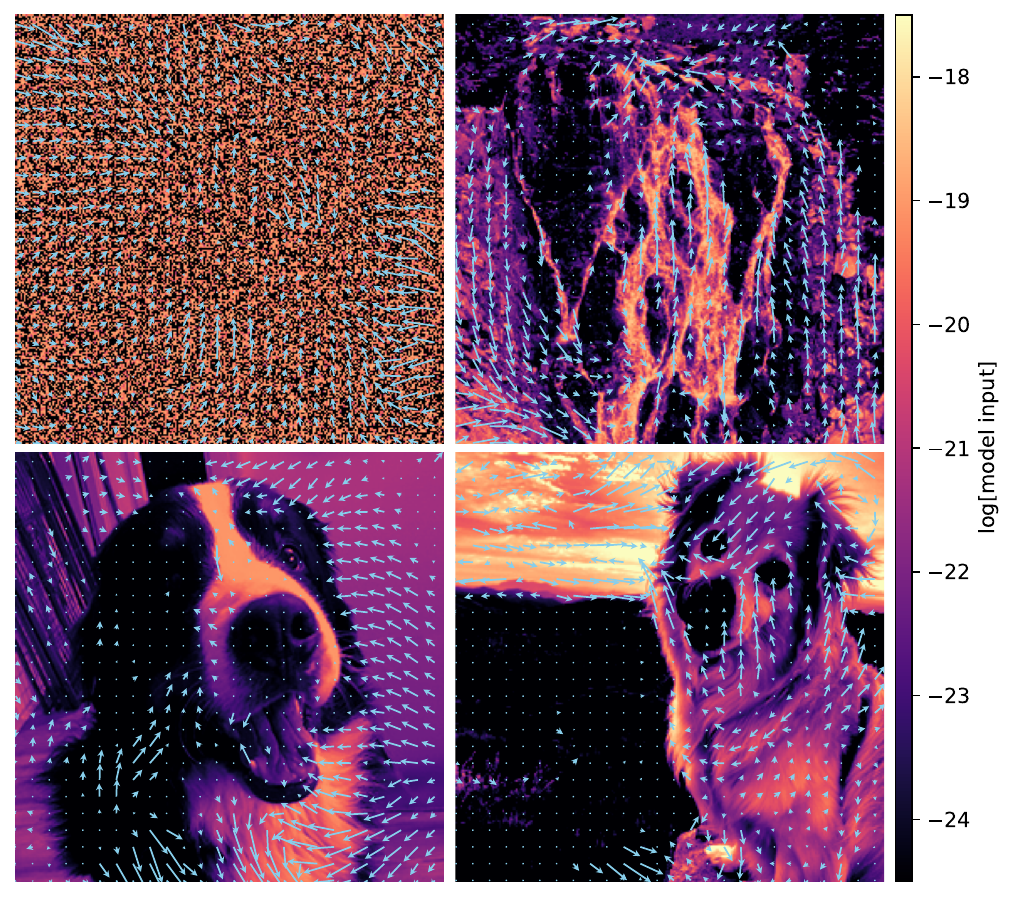}
    \caption{Out-of-distribution test assessing the behavior of the model when inputs are not CGM data. Top Left: output from input of pure random noise, with the same standard deviation and mean as the training data set. Other plots: output from input of structured, non-astronomical data. The colorbar shows the values input into the model; as none of these images have anything to do with H$\alpha$ emission, they are all arbitrarily scaled.}
    \label{fig:OOD3}
\end{figure*}

We experiment with the application of the model to data outside the distribution used for training. Fig.~\ref{fig:OOD} shows the results of applying the model to a massive quiescent galaxy in TNG50, with H$\alpha$ emission and cold gas velocity calculated the same way as for the MW/M31 sample. We deliberately choose a system with plenty of H$\alpha$ emission in the CGM. While there are notable errors, especially in the cloud around the center right and the trail near the top, the model still does a reasonable job predicting the direction and velocity of the inflow on the bottom. This suggests the model will generalize well to galaxies outside of the MW/M31 sample within the TNG50 simulation, especially when such galaxies are included in the training set.

Results of applying the model to non-astronomical data are shown in Fig.~\ref{fig:OOD3}. The first plot shows the results of the model when pure noise is an input. The model infers flows throughout the image, and we re-emphasize that the model is not trained to distinguish regions that have a cold gas signal from regions that do not, only to predict the flow of gas assuming it is present. As mentioned in Sec.~\ref{sec:Discussion}, any application of this model will first require identifying which regions of an image contain cold gas using traditional methods. The remaining input images are photos unrelated to anything in astronomy, scaled arbitrarily to have similar value distributions as the TNG50 emission maps. We caution against reading too much into these results---the model is still a black box---but the results are still interesting for getting some intuition on the features the model cares about. At larger scales, it responds to gradients and the shape of pixels above some brightness level, but it also responds to smaller scale lines and arcs (e.g., the direction of the dogs' fur), which is roughly the expected behavior. The example on the top right of Fig.~\ref{fig:OOD3}, a scaled image of a waterfall, is particularly illustrative since it looks somewhat similar to CGM flows. The model predicts that the water is flowing upwards, possibly due to CGM inflows typically narrowing as they fall towards a galaxy, while the waterfall rivulets start narrow near the top and broaden as they fall.

\bibliography{sample701}{}
\bibliographystyle{aasjournalv7}

\end{document}